\documentclass[aps,twocolumn,prb,showpacs,floatfix]{revtex4}

\usepackage{color}
\usepackage{epsfig,amsmath,amssymb}
\usepackage{graphics} 
\usepackage{subfig}
\usepackage{multirow}

\begin{document}

\title{The frustrated Heisenberg antiferromagnet on the checkerboard lattice:
    \\the $J_{1}$--$J_{2}$ model}

\author{R. F. Bishop and P. H. Y. Li}
\affiliation{School of Physics and Astronomy, Schuster Building, 
The University of Manchester, Manchester M13 9PL, United Kingdom}

\author{D. J. J. Farnell} 
\affiliation{Division of Mathematics, Faculty of Advanced Technology, 
University of Glamorgan, Pontypridd CF37 1DL, Wales, United Kingdom}

\author{J. Richter} 
\affiliation{Institut f\"ur Theoretische Physik, Otto-von-Guericke Universit\"at Magdeburg, P.O.B. 4120, 39016 Magdeburg, Germany}

\author{C.~E.~Campbell}
\affiliation{School of Physics and Astronomy, University of Minnesota, 116 Church Street SE, Minneapolis, Minnesota 55455, USA}

\date{\today}

\begin{abstract}

We study the zero-temperature ground-state (gs) phase diagram of the
spin-$\frac{1}{2}$ anisotropic planar pyrochlore (or crossed
chain) model using the coupled cluster method (CCM).  The model is equivalently 
described as a frustrated $J_{1}$--$J_{2}$ antiferromagnet on the 
two-dimensional checkerboard lattice, with nearest-neighbor exchange bonds of
strength $J_{1}>0$ and next-nearest-neighbor bonds of strength 
$J_{2} \equiv \kappa J_{1} > 0$.  Using various antiferromagnetic (AFM) classical
ground states as CCM model or reference states we present results for the
gs energy, average on-site magnetization, and the susceptibilities of these states
against the formation of plaquette valence-bond crystal (PVBC) 
and crossed-dimer valence-bond crystal (CDVBC) ordering.  Our calculations show 
that the AFM quasiclassical state with N\'{e}el ordering is the gs phase for 
$\kappa < \kappa_{c_1} \approx 0.80 \pm 0.01$, but that none of the fourfold set
of AFM states that are selected by quantum fluctuations at $O(1/s)$ in a large-$s$ 
analysis (where $s$ is the spin quantum number) from the infinitely degenerate 
set of AFM states that form the gs phase for the classical version 
($s \to \infty$) of the model (for $\kappa > 1$) survives the quantum fluctuations 
to form a stable magnetically-ordered gs phase for the $s=\frac{1}{2}$ case.
We show that the quasiclassical N\'{e}el 
state becomes infinitely susceptible to PVBC ordering at or very near to 
$\kappa = \kappa_{c_1}$, and that the quasiclassical fourfold AFM states become
infinitely susceptible to PVBC ordering at $\kappa = \kappa_{c_2} \approx 1.22 \pm 0.02$.
In turn, we find that these states become infinitely susceptible to CDVBC 
ordering for {\it all} values of $\kappa$ above a certain 
critical value at or very near to $\kappa = \kappa_{c_2}$.  Our calculations thus
indicate a N\'{e}el-ordered gs phase for $\kappa < \kappa_{c_1}$, a PVBC-ordered phase
for $\kappa_{c_1} < \kappa < \kappa_{c_2}$, and a CDVBC-ordered  phase 
for $\kappa > \kappa_{c_2}$.  Both transitions are likely to be direct ones, 
although we cannot exclude very narrow coexistence regions confined to 
$0.79 \lesssim \kappa \lesssim 0.81$ and $1.20 \lesssim \kappa \lesssim 1.22$ 
respectively.  
  
\end{abstract}


\maketitle

\section{INTRODUCTION}
The theoretical study of two-dimensional (2D) frustrated quantum antiferromagnets
has been strongly motivated by the fact that such quantum spin models often
describe well the properties of real magnetic materials of great experimental interest.
These models have also become of huge current interest because the interplay
between frustration and quantum fluctuations seen in them can produce, 
even at zero temperature ($T=0$), a wide variety of
fascinating quantum phases ranging from those with quasiclassical
ordering to valence-bond solids and spin 
liquids.\cite{Sachdev:1995,2D_magnetism_1,2D_magnetism_2} They
have thus become paradigms of systems that may be used to study
quantum phase transitions between quasiclassical phases showing magnetic
order and magnetically disordered quantum phases.  

Some of the parameters that determine which 
type of ordering occurs include the lattice geometry, the
dimensionality $D$ of the system, the spin quantum number $s$ of the atoms situated on
the lattice sites, the number and range of the magnetic bonds, and the
degree to which bond frustration of either the geometric or dynamical kind is present. 
New impetus for the study of 2D quantum spin-lattice models 
comes from recent proposals to realize them experimentally with 
ultracold atoms trapped in an optical lattice.\cite{Struck:2011} 
The particularly exciting scenario thus opens of being able 
to tune the competing bond strengths and thus to investigate experimentally
the ensuing quantum phase transitions and their dynamics.

One of the prime theoretical interests in frustrated quantum magnets lies
in the possibility that they might exhibit quantum disordered states and/or
spin-liquid behavior.  Among the most highly frustrated, and hence most
promising, candidate systems in this regard are those that are composed of
tetrahedra coupled into two-dimensional (2D) or three-dimensional (3D) 
lattice networks.  Prominent among the latter are the pyrochlores, whose basic structure
is one of vertex-sharing tetrahedra.  Indeed, experiments on such $s=\frac{1}{2}$
pyrochlores as Y$_{2}$Ir$_{2}$O$_{7}$ do seem to show evidence for a quantum
spin-liquid state.\cite{Fukuzawa:2003}  

\begin{figure*}[!tb]
\mbox{
\subfloat[N\'{e}el]{\scalebox{0.4}{\includegraphics{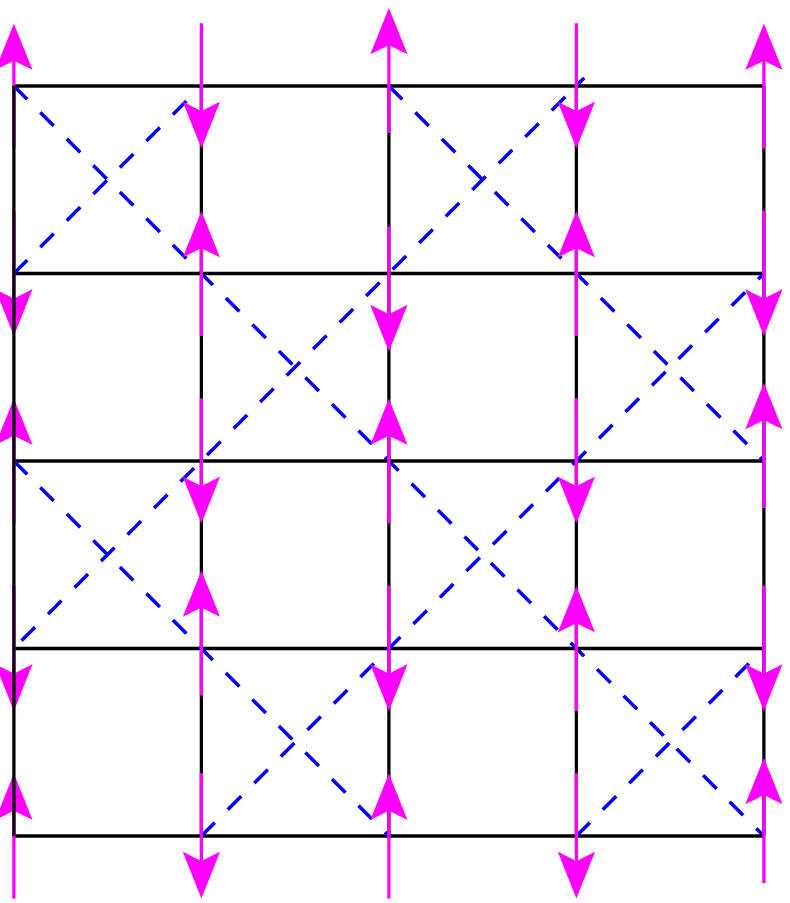}}} \quad \quad
\subfloat[striped]{\scalebox{0.4}{\includegraphics{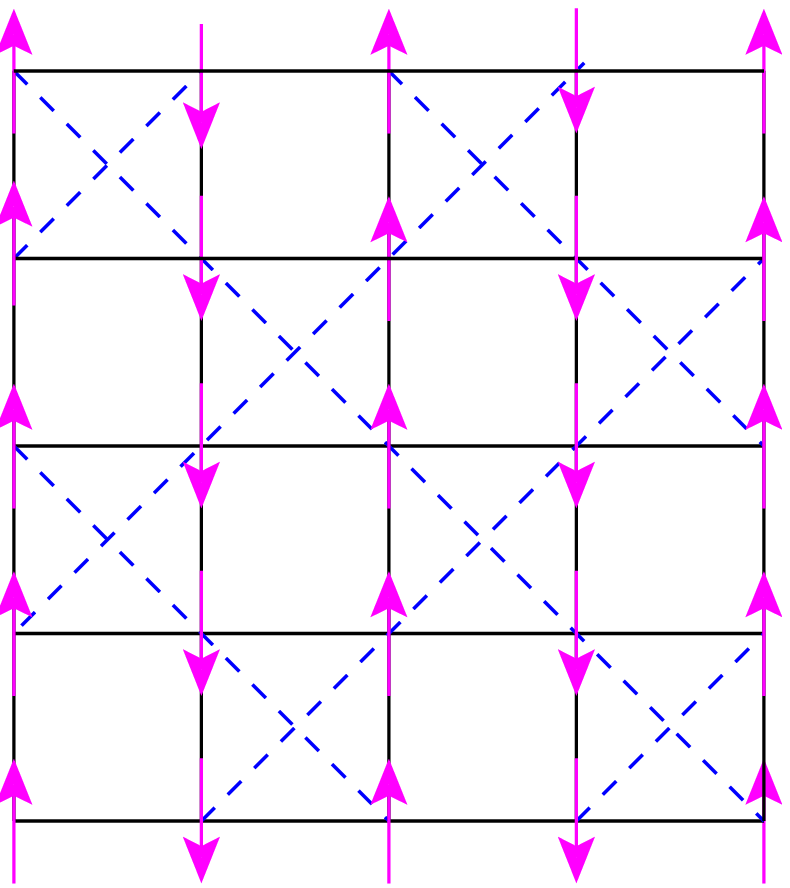}}}  \quad \quad
\subfloat[N\'{e}el$^{\ast}$]{\scalebox{0.4}{\includegraphics{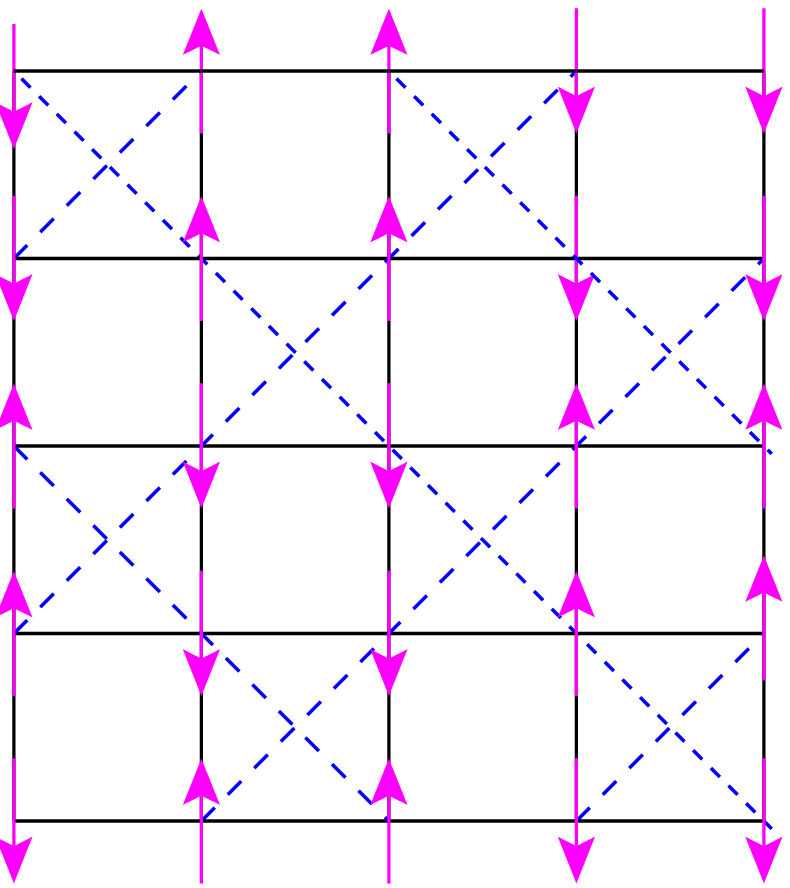}}} 
}
\caption{(Color online) The $J_{1}$--$J_{2}$ checkerboard model (with $J_{1}=1$), showing 
  (a) the N\'{e}el state, (b) the (columnar) striped state and (c) one of the two N\'{e}el$^{\ast}$ states.
  The NN $J_{1}$ bonds are shown as solid (black) lines and the NNN 
  $J_{2}$ bonds are shown as dashed (blue) lines.  The arrows
  represent the orientations of the spins on each lattice site 
  for each of the three states shown.}
\label{model_bonds}
\end{figure*}

In order to reduce the complexity
of the 3D pyrochlore lattice, but without diminishing the magnetic frustration,
one may project the 3D vertex-sharing lattice of tetrahedra onto a 2D plane.  Each
tetrahedron comprises four spins at its vertices, with each of its six edges or links
representing an interaction of the Heisenberg antiferromagnet (HAFM) form.  
Each such tetrahedron is thus mapped to a square with spins at its vertices and with sides
representing antiferromagnetic (AFM) bonds, but now with additional AFM links across 
its diagonals.  Such a pattern is repeated in the vertex-sharing arrangement shown in
the checkerboard pattern of Fig.~\ref{model_bonds}.
Although this 2D projection of the 3D pyrochlore structure preserves its vertex-sharing
structure, the symmetry in the 3D structure between the six bonds on each tetrahedron is 
lost in the 2D projection since the two diagonal bonds of each crossed square are
now inequivalent to the four bonds on the sides of the square.  This subsequent reduction in
the symmetry is thus consistent with considering an anisotropic Heisenberg model on 
the 2D checkerboard lattice in which the AFM exchange interactions along the sides
of the squares (with strength $J_{1}>0$) are generally different in strength from 
those along the diagonals of the crossed squares (which have a strength $J_{2}>0$),
as shown in Fig.~\ref{model_bonds}.  The resulting frustrated model is thus called
the anisotropic planar pyrochlore.  Alternative names are the anisotropic checkerboard 
HAFM, the $J_{1}$--$J_{2}$ checkerboard model, and the crossed chain model.

Although, the spin-$\frac{1}{2}$ anisotropic planar pyrochlore has been studied by a 
large number of authors\cite{Singh:1998,Palmer:2001,Chung:2001,Brenig:2002,Canals:2002,
Starykh:2002,Sindzingre:2002,Fouet:2003,Berg:2003,Tchernyshyov:2003,Moessner:2004,Hermele:2004,
Brenig:2004,Bernier:2004,Starykh:2005,Schmidt:2006,Arlego:2007,Moukouri:2008,Chan:2011} 
the structure of its full phase diagram still remains
unsettled and contentious, especially for larger values of the frustration parameter, 
$\kappa \equiv J_{2}/J_{1} \gtrsim 1$, as we discuss more fully below in Sec.~\ref{model}.  
Various methods have been applied to the model for different regions of the parameter space
for the variable $\kappa$.  These include semiclassical ($s \gg 1$)
analyses,\cite{Singh:1998,Canals:2002,Tchernyshyov:2003} large-$N$
expansions of the Sp($N$) model,\cite{Chung:2001,Moessner:2004,Bernier:2004} high-order
cluster-based strong-coupling series expansion (SE)  techniques\cite{Brenig:2002,Brenig:2004,Arlego:2007} 
using a continuous unitary transformation generated by the flow equation method of Wegner,\cite{Wegner:1994} 
a real-space renormalization technique\cite{Berg:2003} using
the contractor renormalization method of Morningstar and Weinstein,\cite{Morningstar:1996}
an easy-axis generalization of the 3D model,\cite{Hermele:2004} a quasi-one-dimensional
approach (valid in the $\kappa \gg 1$ limit) based on the random phase approximation
backed up by a bosonization study,\cite{Starykh:2002} techniques that
combine renormalization group ideas with one-dimensional bosonization and current
algebra methods,\cite{Starykh:2005} exact diagonalization (ED) of small finite-lattice 
clusters,\cite{Palmer:2001,Sindzingre:2002,Fouet:2003,Schmidt:2006} a two-step 
density-matrix renormalization group method,\cite{Moukouri:2008} and, very recently,
a tensor network simulation\cite{Chan:2011} based on infinite projected entangled pair states.\cite{Jordan:2008}

In this paper we use the coupled cluster method (CCM) of quantum many-body theory
(see, e.g., Refs.~[\onlinecite{Bi:1991,Bi:1998,Fa:2004}] and references cited therein) to study 
the spin-$\frac{1}{2}$ $J_{1}$--$J_{2}$ Heisenberg model on the checkerboard
lattice, in order to attempt to shed more light on it.  The CCM has proven itself in many
applications to frustrated magnetic systems to be capable of providing accurate estimates
of the quantum critical points marking the phase transitions between states of widely differing order (see e.g.,
Refs.~[\onlinecite{Ze:1998,Kr:2000,ccm3,Fa:2001,schmalfuss,Fa:2004,rachid05,darradi08,Bi:2008_JPCM,
Bi:2008_PRB,richter10,UJack_ccm,Reuther:2011_J1J2J3mod,Farnell:2011,Gotze:2011}]).  
In view of the continuing interest in the model and
the controversy over its $T=0$ phase structure, especially at large frustration ($\kappa \gtrsim 1$),
it seems appropriate and timely to bring to bear on the problem the proven power of the CCM.  Since,
as we shall see, we are able to calculate to high orders in the relevant CCM approximation
scheme, as discussed below in Sec.~\ref{CCM}, we are able to present accurate results
from a method of well-proven ability to deal with such strongly correlated
and highly frustrated systems.

We now briefly outline the structure of the remainder of the paper.  In Sec.~\ref{model} the
model itself is first described and discussed.  The CCM formalism is then briefly outlined in
Sec.~\ref{CCM} before we present and discuss our CCM results in
Sec.~\ref{results}.  Finally, we conclude in Sec.~\ref{summary} with a
summary of our findings and a comparison of our results for the model
with those from other methods.

\section{THE MODEL}
\label{model}
The Hamiltonian for the anisotropic checkerboard-lattice model considered here is given by
\begin{equation}
H = J_{1}\sum_{\langle i,j \rangle} \mathbf{s}_{i}\cdot\mathbf{s}_{j} + J_{2}\sum_{\langle\langle i,k \rangle\rangle} 
\mathbf{s}_{i}\cdot\mathbf{s}_{k}\,. \label{H}
\end{equation}
where the index $i$ runs over all sites of a square lattice, 
index $j$ runs over all nearest-neighbor (NN) sites to site $i$, and
index $k$ runs over all next-nearest-neighbor (NNN) sites to site $i$ on
a checkerboard pattern such that alternate square plaquettes have either two 
NNN (diagonal) bonds or none, as shown in Fig.~\ref{model_bonds}. 
The sums over $\langle i,j \rangle$ and
$\langle \langle i,k \rangle \rangle$ count each pairwise bond
once and once only.  Each site $i$ of the lattice carries a particle with
spin $s=\frac{1}{2}$ and a spin operator 
${\bf s}_{i}=(s_{i}^{x},s_{i}^{y},s_{i}^{z})$.   

The lattice and exchange bonds of the anisotropic checkerboard-lattice model are shown in Fig.~\ref{model_bonds}.
We may alternatively view the model as comprising crossed (diagonal) 
sets of chains on which the intrachain exchange coupling constant is
$J_2$, coupled by (vertical and horizontal) interchain exchange bonds
of strength $J_1$.  We assume here that both bonds are antiferromagnetic 
(AFM) in nature (i.e., have positive exchange coupling constants) and hence
frustrate one another.  The model thus interpolates continuously between the isotropic HAFM 
on the square lattice (when $\kappa \equiv J_{2}/J_{1}=0$) and decoupled
one-dimensional (1D) isotropic HAFM chains (when $\kappa \to \infty$).  In between, at
$\kappa = 1$, we have the isotropic HAFM on the checkerboard lattice that is
a 2D analog of the 3D isotropic pyrochlore HAFM.  Henceforth, without 
loss of generality, we set $J_{1} \equiv 1$ in order to set the energy scale.

The classical ground-state (gs) phase for this model for $\kappa < 1$
is the N\'e{e}l state shown in Fig.~\ref{model_bonds}(a), in which every column and row exhibits
N\'e{e}l AFM ordering, $\cdots \uparrow \downarrow \uparrow \downarrow \cdots$, 
and consequently the ordering along each diagonal is
ferromagnetic (FM), i.e., where all the spins are aligned parallel to
one another.  The N\'e{e}l state has an energy per spin given by
$E^{\rm cl}/N=s^{2}(-2J_{1}+J_{2})$.
For $\kappa > 1$ there is an
infinitely degenerate family of collinear gs phases in which every
diagonal exhibits N\'e{e}l AFM ordering, but where every diagonal, each of which is 
connected by $J_1$ bonds to two other crossed diagonals, 
can be arbitrarily moved along its own direction.  These states all 
have the same energy per spin of $E^{\rm cl}/N=-s^{2}J_{2}$,
independent of the exchange coupling $J_1$.  The classical phase transition is clearly 
at $\kappa_{{\rm cl}}=1$ ($J_{1}>0$).  

Among this infinitely degenerate family of classical states for $\kappa > 1$ are the so-called
(columnar) striped state shown in Fig.~\ref{model_bonds}(b) and 
the N\'{e}el$^{\ast}$ state shown in Fig.~\ref{model_bonds}(c).  The columnar (row) striped states
have FM ordering along columns (rows) but AFM N\'e{e}l ordering along rows (columns).
The N\'{e}el$^{\ast}$ state has doubled AFM ordering, 
$\cdots \uparrow \uparrow \downarrow \downarrow \uparrow \uparrow \downarrow \downarrow \cdots$,
along every row and column.  Thus, the single-site spin $\uparrow$ or $\downarrow$ of the
usual N\'{e}el state is replaced in the N\'{e}el$^{\ast}$ state by the two-site unit 
$\uparrow\uparrow$ or $\downarrow\downarrow$.
Like the striped state, the N\'{e}el$^{\ast}$ state is also 
doubly degenerate (for a given direction of the N\'{e}el vector), 
since the roles of the rows and columns may be interchanged in Fig.~\ref{model_bonds}(c) (or,
equivalently, the two-site unit of up or down spins may be chosen along rows as well as columns). 

Compared to the classical ($s \to \infty$) version of the anisotropic checkerboard model,
the $s = \frac{1}{2}$ case is really only well established at the three points $\kappa = 0$, 
$\kappa = 1$, and $\kappa \to \infty$.  For the square-lattice HAFM ($\kappa = 0$) 
almost all methods concur that the classical N\'{e}el AFM long-range order (LRO) 
is not destroyed, although the staggered magnetization is reduced from the 
classical value of 0.5, and the excitations are gapless, integer-spin magnons.
By continuity it is expected that the N\'{e}el order 
will persist as the frustrating $J_2$-bonds are turned on, out to some critical
value $\kappa_{c_1}$, at which the N\'{e}el staggered magnetization goes to zero.  

There is also a broad general consensus from a variety of methods that at the isotropic
point ($\kappa = 1$) the gs phase of the $s = \frac{1}{2}$ checkerboard-lattice HAFM is a
plaquette valence-bond crystal (PVBC) with quadrumer LRO on isolated spin-singlet 
square plaquettes, and with gapped integer-spin excitations 
(that are confined spinons).  It is still an open question as
to whether there is a direct (first-order in the Landau-Ginzburg scenario) transition at $\kappa = \kappa_{c_1}$
between the states with N\'{e}el and PVBC order, or whether there is an
intermediate coexistence phase with two different order parameters.  Such a phase
could have continuous Landau-Ginzburg transitions to both the 
N\'{e}el and PVBC phases.  The possibility of such coexistence regions occurring between
N\'{e}el and valence-bond solids has been discussed in great detail both in a general context in
Ref.~[\onlinecite{Sachdev:2002}] for various spin-lattice models, and in 
Ref.~[\onlinecite{Starykh:2005}] in the specific context of the present model.  Again, by continuity,
we expect that the PVBC order will persist to values of $\kappa$ out to some critical
value $\kappa_{c_2}>1$, at which point the PVBC order vanishes.

Lastly, at the $\kappa \to \infty$ limit of the $s = \frac{1}{2}$ anisotropic checkerboard 
model we have the well-known and exactly soluble case of decoupled 1D HAFM chains.  Such 1D
spin-$\frac{1}{2}$ chains have a Luttinger spin-liquid gs phase, with a gapless
excitation spectrum of deconfined spin-$\frac{1}{2}$ spinons.  

The most unsettled part of the phase diagram for this model is the region 
$\kappa \gtrsim \kappa_{c_2}$, where various predictions have been given.  For example,
it has been argued\cite{Starykh:2002} that the 1D Luttinger behavior of the $\kappa \to \infty$
limit might be robust against the turning on of interchain ($J_1$) couplings, so that
the chains continue to act as decoupled.  Such a 2D spin-liquid gs phase provides an
example of a so-called sliding Luttinger liquid.\cite{Emery:2000,
Mukhopadhyay:2001,Vishwanath:2001}  Numerical evidence for such a spin-liquid
phase at large values of $\kappa$ in the present model was also found 
from ED studies on samples of up to $N=36$ spins.\cite{Sindzingre:2002}  

Alternatively, by making a more careful analysis of the relevant terms
near the 1D Luttinger liquid fixed point, it was shown later\cite{Starykh:2005} that
the original prediction\cite{Starykh:2002} of a sliding Luttinger liquid was 
wrong, and the same authors suggested that the correct gs phase in the 
large-$\kappa$ limit is the so-called gapped crossed dimer phase, where the system
spontaneously dimerizes with a staggered ordering of dimers along the 
$J_2$ chains (i.e., along the diagonals in Fig.~\ref{model_bonds}).  Support
for the crossed-dimer phase has come from series-expansion\cite{Arlego:2007} and 
two-step density-matrix renormalization group method studies.\cite{Moukouri:2008}  
We discuss this phase further in Sec.~\ref{results} below.

Finally, one may wonder whether any of the infinitely-degenerate set of classical
($s \to \infty$) ground states for $\kappa > 1$ may survive the quantum
fluctuations present in the $s = \frac{1}{2}$ model, and, if so, whether the 
classical degeneracy may be lifted by the well-known {\it order by disorder}
mechanism.\cite{Villain:1977}  A semiclassical ($s \gg 1$) 
analysis\cite{Tchernyshyov:2003} has shown that quantum spin-fluctuations induce
a LRO that breaks the fourfold rotational symmetry of the lattice, and
that to $O(1/s)$ the fourfold degenerate set of states comprising the 
striped state of Fig.~\ref{model_bonds}(b) and the N\'{e}el$^{\ast}$ state 
of Fig.~\ref{model_bonds}(c) (plus their two counterparts where rows and
columns are interchanged) become energetically favored as the gs phase over
the remainder of the infinite classical set.  A very recent tensor network
simulation\cite{Chan:2011} of the spin-$\frac{1}{2}$ model finds that,
contrary to essentially all other calculations on this model, this fourfold
degenerate state survives to be the quantum gs phase for all values of
the frustration parameter above that at which PVBC order disappears
($\kappa > \kappa_{c_2}$).  These authors also argue that, although their
numerical program is unable to distinguish between the energies of
the striped and N\'{e}el$^{\ast}$ states in the quantum ($s=\frac{1}{2}$) model,
the striped phase will emerge as the actual gs phase in practice because of 
its greater robustness against small perturbations to the Hamiltonian.

Other analyses\cite{Starykh:2005} have, however, shown that, the N\'{e}el$^{\ast}$ state
might, in one possible scenario, intervene as an intermediate gs phase between the 
two (i.e., the plaquette and crossed-dimer) valence-bond solid phases.  In such a scenario
the transition between the PVBC and N\'{e}el$^{\ast}$ phases is shown to be able
to proceed via a continuous O(3) transition, while that between the crossed dimer and 
N\'{e}el$^{\ast}$ phases will be either a direct (first-order in the Landau-Ginzburg scenario) one or
will proceed via an intermediate coexistence phase showing both types of ordering (i.e., both
N\'{e}el$^{\ast}$ spin ordering and crossed dimer bond modulation).

In view of the considerable lack of agreement about the gs phase diagram for the 
spin-$\frac{1}{2}$ $J_{1}$--$J_{2}$ model on the checkerboard lattice we now present
results for it in the present paper from high-order CCM calculations.

\section{THE COUPLED CLUSTER METHOD}
\label{CCM}
The CCM (see, e.g., Refs.~[\onlinecite{Bi:1991,Bi:1998,Fa:2004}] and
references cited therein) is one of the most powerful and universally applicable
quantum many-body techniques.  It has been applied successfully to many
quantum spin-systems (see e.g.,
Refs.~[\onlinecite{Ze:1998,Kr:2000,ccm3,Fa:2001,schmalfuss,Fa:2004,rachid05,darradi08,Bi:2008_JPCM,
Bi:2008_PRB,richter10,UJack_ccm,Reuther:2011_J1J2J3mod,Farnell:2011,Gotze:2011}]
and references cited therein).  The method is particularly suitable
for investigating highly frustrated quantum magnets, for which
other alternative methods may be of limited usefulness.  For instance, quantum
Monte Carlo (QMC) techniques are often severely restricted by the 
well-known ``minus-sign problem,'' which is
ubiquitous in highly frustrated quantum magnets.  On the other hand,
ED methods are often too restricted by the relatively small size of
the largest lattices that can be handled with given computational resources to
be able to sample accurately the often subtle ordering present.

We briefly describe the CCM formalism here and we refer the
interested reader to the literature (and see, e.g., 
Refs.~[\onlinecite{Ze:1998,Kr:2000,ccm3,Fa:2001,schmalfuss,Fa:2004,rachid05,darradi08,Bi:2008_JPCM,Bi:2008_PRB}]
and references cited therein) for further details.  The implementation of the CCM always
begins with the choice of a suitable reference or model state.  It is usual, but 
by no means vital, to choose a classical gs phase as the
model state $|\Phi\rangle$.  Hence, for the present anisotropic checkerboard model, we choose the
N\'{e}el state, the striped state and the N\'{e}el$^{\ast}$ state as our CCM
model states.  From the discussion in Sec.~\ref{model} above we expect that the 
N\'{e}el state is likely to provide a good candidate CCM model state in the region
$\kappa \lesssim 1$, while the striped and the N\'{e}el$^{\ast}$ states are expected
to be suitable candidates for $\kappa \gtrsim 1$.  We choose only the latter states out
of the infinitely degenerate set of classical states in the $\kappa > 1$ regime since,
as discussed above, this fourfold set of states is selected by the order by disorder
mechanism at the $O(1/s)$ level in a quasiclassical expansion in powers of 
$1/s$,\cite{Tchernyshyov:2003} at which order they remain degenerate in energy.

The CCM then incorporates the multi-particle correlations present 
in the exact quantum gs phase under investigation on top of the chosen model state in a 
systematic hierarchy of approximations for the correlation operators $S$ 
and $\tilde{S}$ which parametrize the gs ket and bra wave functions as 
\begin{equation}
|\Psi\rangle=e^{S}|\Phi\rangle; \quad \langle \tilde{\Psi}|=\langle
\Phi|\tilde{S}e^{-S}. 
\label{psi}
\end{equation}
The correlation operators are written as
\begin{equation}
S=\sum_{I\neq0}{\cal S}_{I}C^{+}_{I}; \quad \tilde{S}=\sum_{I\neq0}\tilde{\cal S}_{I}C^{-}_{I}; \quad \forall I \neq 0,
\label{S}
\end{equation}
where $C^{+}_{0} \equiv 1$, the identity operator, 
$I$ is a set-index describing a set of single-particle configurations, and 
$C^{+}_{I}$ and $C^{-}_{I} \equiv (C^{+}_{I})^{\dagger}$, for $I \neq 0$, are 
Hermitian-conjugate pairs of multi-particle creation and
destruction operators defined with respect to the model state
$|\Phi\rangle$ considered as a generalized vacuum state. 
They are thus required to satisfy the conditions $\langle \Phi |C^{+}_{I} =
0 = C^{-}_{I}| \Phi \rangle; \forall I \neq 0$.
They form a complete set of mutually commuting many-body creation operators 
in the Hilbert space, defined with respect to $|\Phi\rangle$ as a cyclic vector. 
The states are normalized such that $\langle\tilde{\Psi}|\Psi\rangle =
\langle \Phi| \Psi\rangle = \langle \Phi| \Phi \rangle \equiv 1$.

For spin-lattice systems it is convenient to choose a set of local
coordinate frames in spin space such that on each lattice site
the spin in each model state points in the downward (negative $z$
direction).  Such rotations obviously do not affect the basic
SU(2) spin commutation relations, but they have the simplifying effect
that the operators $C^{+}_{I}$
are transformed into multi-spin raising operators that can be
expressed as products of single-spin raising operators, $C^{+}_{I}\equiv s^{+}_{j_{1}}
s^{+}_{j_{2}} \cdots s^{+}_{j_{n}}$, where $s^{+}_{j} \equiv s^{x}_{j} + is^{y}_{j}$.  

The gs energy is evaluated in terms of the correlation 
coefficients $\{{\cal S}_{I}\}$ as $E=\langle\tilde{\Psi}|H|\Psi\rangle = 
\langle\Phi|\mbox{e}^{-S}H\mbox{e}^{S}|\Phi\rangle$; and the average on-site
magnetization $M$ in the rotated spin coordinates is evaluated
equivalently in terms of the coefficients $\{{\cal S}_{I},\tilde{{\cal S}_{I}}\}$ 
as $M \equiv -\frac{1}{N} \langle\tilde{\Psi}|\sum_{j=1}^{N}s^{z}_{j}|\Psi\rangle$. 
Thus, $M$ is simply the usual magnetic order parameter.

The complete set of unknown ket- and bra-state correlation coefficients 
$\{{\cal S}_{I}, \tilde{{\cal S}_{I}}\}$ is evaluated by setting the energy expectation
value $\bar{H} \equiv \langle\tilde{\Psi}|H|\Psi\rangle$ to be a minimum with
respect to all parameters $\{{\cal S}_{I}, \tilde{{\cal S}_{I}}; \forall I \neq 0\}$.
This produces the coupled set of nonlinear equations for the ket-state (creation) 
correlation coefficients $\{{\cal S}_{I}\}$ via $\langle \Phi|C^{-}_{I}\mbox{e}^{-S}H\mbox{e}^{S}|\Phi\rangle =
0; \forall I \neq 0$; plus the coupled set of linear equations, 
$\langle\Phi|\tilde{S}(\mbox{e}^{-S}H\mbox{e}^{S} - E)C^{+}_{I}|\Phi\rangle = 0; \forall I \neq 0$, 
which are used to solve for the bra-state (destruction) correlation coefficients
$\{\tilde{{\cal S}_{I}}\}$.  

If it were possible to consider all creation and annihilation operators $C^{+}_{I}$
and  $C^{-}_{I}$ respectively, i.e., all sets (configurations) of lattice sites, in
the CCM correlation operators $S$ and $\tilde{S}$ respectively, one would
in principle obtain the exact eigenstate of the system belonging to
any symmetries imposed by the model state (and the configurations 
that are perhaps also accordingly selected).\cite{Bi:1998} 
Of course, however, it is necessary in practice to use approximations schemes
to truncate the expansions of $S$ and $\tilde{S}$ in Eq.~(\ref{S}).  In that case
the approximate results for the gs energy $E$ and the magnetization $M$ will
depend on the choice of model state.

For the case of $s=\frac{1}{2}$ systems, 
as considered here, we normally use the well-tested
localized LSUB$m$ truncation scheme which takes in 
at the $m$th level of approximation all multi-spin
correlations in the CCM correlation operators over all configured regions on the
lattice defined by $m$ or fewer contiguous sites.  A configuration
of $m$ sites is considered to be contiguous if every site 
in the configuration is adjacent (in the NN sense) to at least one 
other site in the configuration.  Clearly, as $m \to \infty$, the
LSUB$m$ approximation becomes exact.  For the present
checkerboard model, we use the CCM and the LSUB$m$ scheme with $m \leq 10$ for
the three model states shown in Fig.~\ref{model_bonds}.  For the LSUB$m$
configurations we assume the fundamental checkerboard geometry to
define the LSUB$m$ sequences, and hence treat both the pairs of  
sites connected by $J_{1}$ bonds and those connected by
$J_{2}$ bonds as being contiguous sites.  
Table~\ref{table_FundConfig} shows the number $N_{f}$ of such distinct
(i.e., under the symmetries of the lattice and the model state)
fundamental spin configurations for each of the three model states that we
use for our spin-$\frac{1}{2}$ $J_{1}$--$J_{2}$ checkerboard model.
\begin{table}
  \caption{Number of fundamental configurations, $N_{f}$, for the checkerboard geometry for the 
spin-$\frac{1}{2}$ $J_{1}$--$J_{2}$ checkerboard model ($J_{1}=1$), 
using the N\'{e}el, striped, and N\'{e}el$^{\ast}$ states as CCM model states.}
\vskip0.2cm
\begin{tabular}{|c|c|c|c|} \hline\hline
Method & \multicolumn{3}{|c|}{$N_{f}$} \\ \cline{2-4}    
&      N\'{e}el & striped  &  N\'{e}el$^{\ast}$  \\ \hline
LSUB4 & 27       & 54      & 79     \\
LSUB6 & 632     & 1225      & 2441     \\ 
LSUB8 & 21317     & 41324     & 86590   \\
LSUB10 & 825851    & 1598675 &  3373495 \\   \hline\hline     
\end{tabular} 
\label{table_FundConfig}
\end{table}

It is clear that $N_{f}$ rises rapidly with the truncation index $m$.  For example, 
for the N\'{e}el$^{\ast}$ state that is used in this study as one of the CCM model states, 
the LSUB10 approximation contains 3373495 distinct spin
configurations.  This is the highest LSUB$m$ level that we can reach here using 
the N\'{e}el$^{\ast}$ state as our model state, even
with massive parallelization and the use of supercomputing resources.  
It takes us approximately 1~h computing time using
massively parallel computing with 3000 processors simultaneously to
solve the corresponding coupled sets of CCM bra- and ket-state
equations, to obtain a single data point for a given value of $J_{2}$,
with $J_{1}=1$.\cite{ccm}

We note that if, instead of using the checkerboard geometry, we were to use the
square-lattice geometry (i.e., with NN pairs defined only by $J_{1}$ bonds),
the number of fundamental configurations $N_{f}$ would
obviously be fewer, at the same level $m$, than in the checkerboard geometry.
In turn this could perhaps enable us go to higher LSUB$m$
orders for given computational power.  However, this advantage is completely outweighed by the
disadvantage that the LSUB$m$ sequences for both $E/N$ and $M$ then
show a marked staggering behavior in $m \equiv 2k$, depending on whether $k$ 
is even or odd.  This is clearly due to the fact that the full LSUB$m$ sequence does not 
then properly reflect the checkerboard symmetries.  It is quite similar 
to the odd and even staggering behavior in index $m$ for LSUB$m$ approximations 
on simple (dynamically unfrustrated) models, which has been reported
elsewhere.\cite{Farnell:2008}  Any such staggering effect makes
extrapolations (for the full sequence) of the sort we now discuss more 
complicated and less robust.

\begin{figure*}[!tb]
\mbox{
   \subfloat[$E/N$]{\scalebox{0.3}{\includegraphics[angle=270]{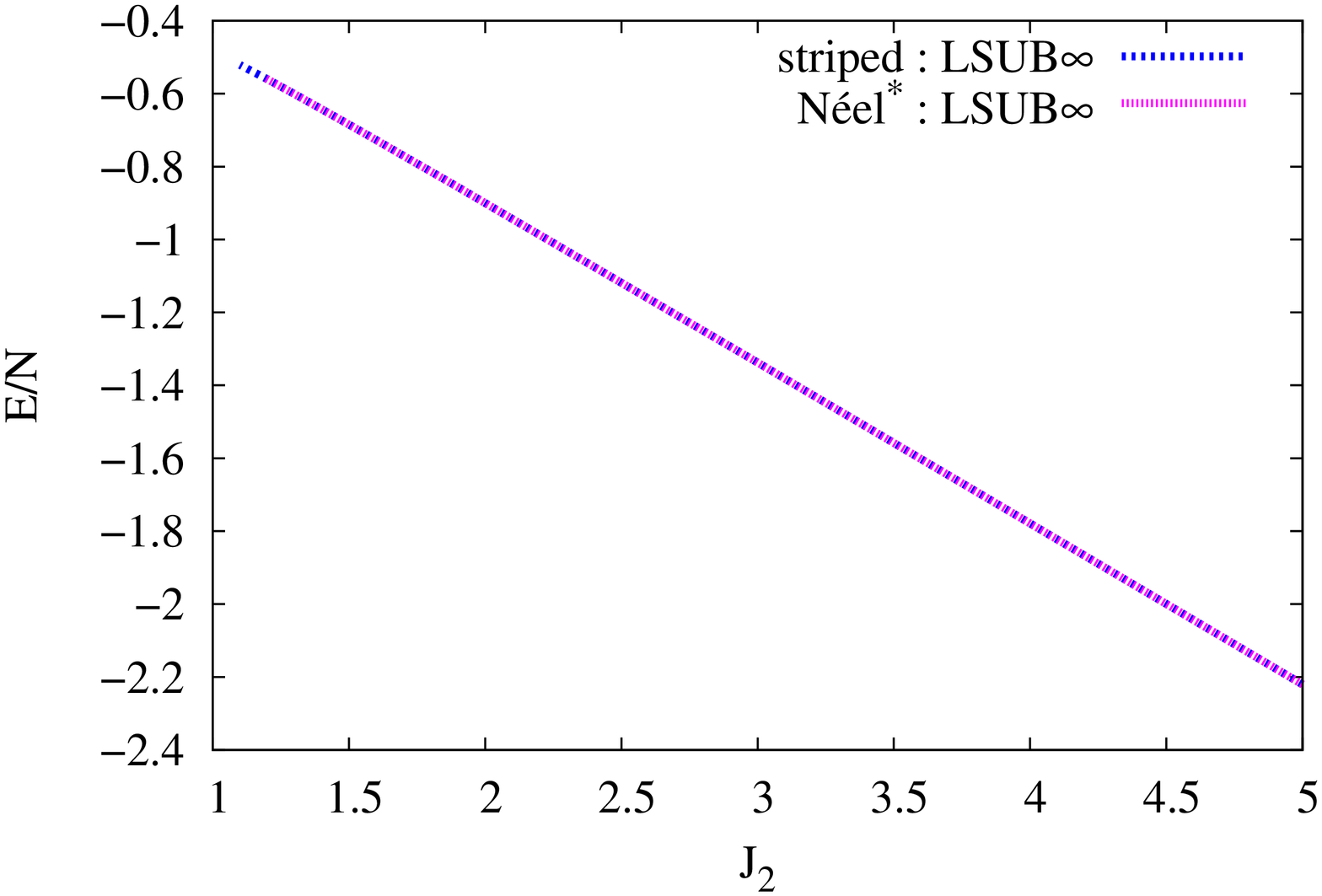}}}
\subfloat[$\Delta e$]{\scalebox{0.3}{\includegraphics[angle=270]{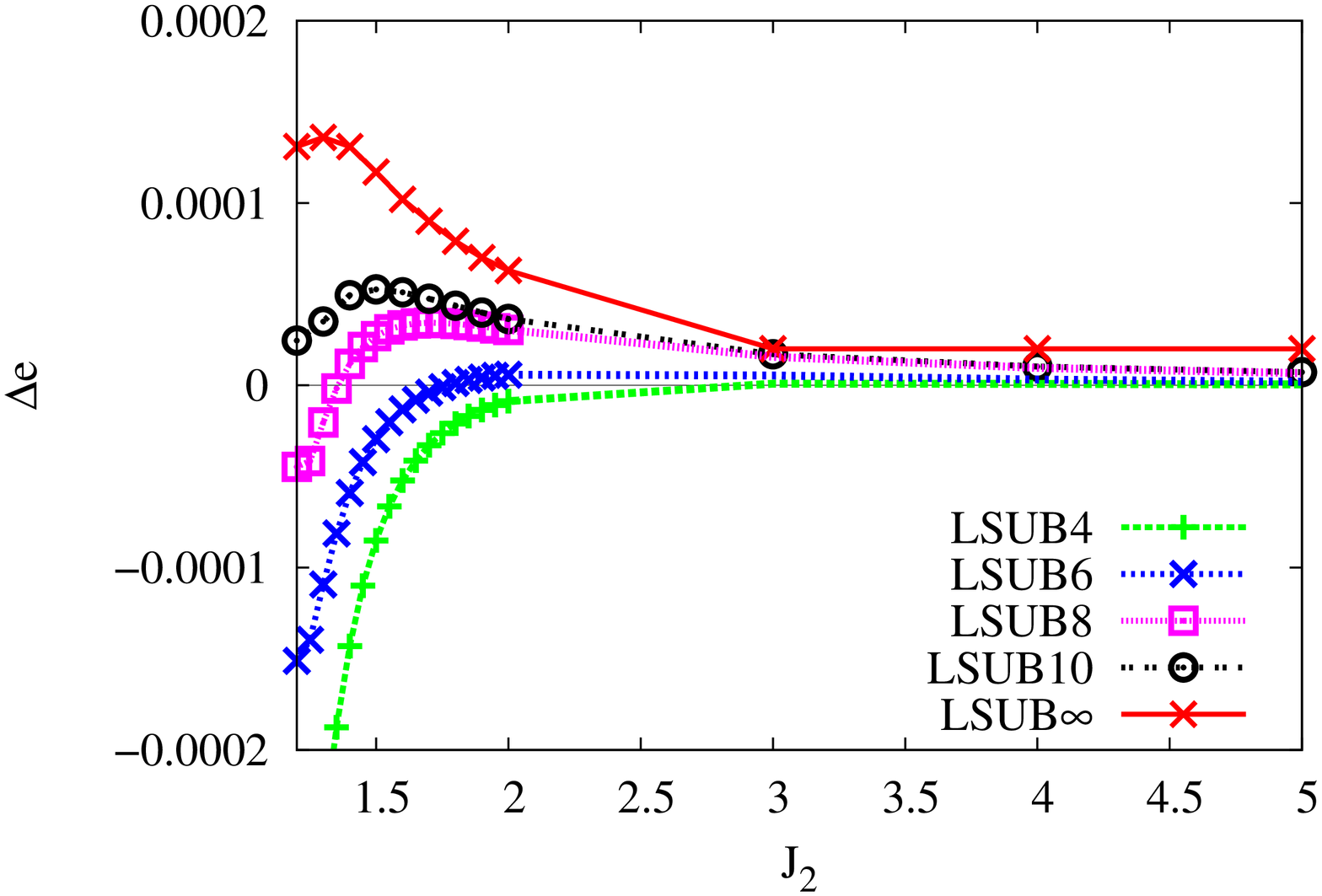}}}
}
\caption{(a) The extrapolated CCM LSUB$\infty$ results for the gs energy 
  per spin, $e \equiv E/N$, versus $J_2$ for the striped and N\'{e}el$^{\ast}$ phases of the 
  spin-$\frac{1}{2}$ $J_{1}$--$J_{2}$ Heisenberg antiferromagnet on
  the checkerboard lattice (with $J_{1}=1$), using the LSUB$m$
  results with $m=\{4,6,8,10\}$ and Eq.~(\ref{Extrapo_E}); (b) the energy
  difference, $\Delta e \equiv e^{{\rm striped}} - e^{{\rm N\acute{e}el}^{\ast}}$ 
  versus $J_{2}$ of the two phases shown in (a) using LSUB$m$ approximations
  with $m=\{4,6,8,10\}$ and also using the corresponding separate LSUB$\infty$ results
  for both phases from Eq.~(\ref{Extrapo_E}) using $m=\{4,6,8,10\}$.}
\label{Ediff_ColStriped_stateC}
\end{figure*}
 
Thus, as a final step we need to extrapolate the raw CCM data from our LSUB$m$ 
approximations to the exact ($m \to \infty$) limit.  In the absence 
of any staggering effects of the sort described above, we use the well-tested 
extrapolation scheme
\begin{equation}
E(m)/N=a_{0}+a_{1}m^{-2}+a_{2}m^{-4}\,,             \label{Extrapo_E}
\end{equation}
for the gs energy.\cite{Kr:2000,ccm3,Fa:2001,rachid05,schmalfuss,Bi:2008_PRB,darradi08,
Bi:2008_JPCM,richter10,Reuther:2011_J1J2J3mod}  
For the magnetic order parameter, $M$, we use the schemes
\begin{equation}
M(m)=b_{0}+b_{1}m^{-1}+b_{2}m^{-2}\,,             \label{Extrapo_M}
\end{equation}
for non-frustrated spin systems,\cite{Kr:2000,ccm3,Fa:2001} and
\begin{equation}
M(m)=c_{0}+b_{1}m^{-1/2}+a_{2}m^{-3/2}\,,             \label{Extrapo_M_Frust}
\end{equation}
for highly frustrated spin systems.\cite{darradi08,Bi:2008_JPCM,richter10,Reuther:2011_J1J2J3mod}  
We have performed separate extrapolations using data 
sets with $m=\{4,6,8,10\}$, $m=\{6,8,10\}$, $m=\{2,4,6,8\}$, and $m=\{4,6,8\}$.  
They yield very similar results in each of the cases reported below, 
which gives credence to our results and demonstrates their robustness.

\section{RESULTS AND DISCUSSION}
\label{results}
We now present our CCM results for the spin-$\frac{1}{2}$ anisotropic
checkerboard model, using each of the three states shown in 
Fig.~\ref{model_bonds} as model states.  We first show in 
Fig.~\ref{Ediff_ColStriped_stateC}(a) the extrapolated LSUB$\infty$ gs energies per spin,
$E/N$, of the phases obtained using the striped and N\'{e}el$^{\ast}$ 
model states.  We recall that in the classical limit ($s \rightarrow \infty$)
these two phases are degenerate and are the gs phase only for $\kappa > 1$.  
Results are shown in Fig.~\ref{Ediff_ColStriped_stateC}(a) down to the lowest 
terminating values of $\kappa$ in each case for which real solutions exist
for all of the LSUB$m$ approximations used.  It has been shown 
previously\cite{Fa:2004,UJack_ccm} that such termination points are
a strong indication of the corresponding quantum phase transition points
that occur in the system under study.
Figure \ref{Ediff_ColStriped_stateC}(b) shows the difference in the energies
of the two states in the approximate region where both CCM solutions exist.
We see clearly that although the energy difference is small, the classical
degeneracy is removed in favor of the N\'{e}el$^{\ast}$ state over the
striped state for all values of $\kappa$ for which real CCM solutions for
both phases exist.  Nevertheless, bearing in mind the smallness of 
the energy difference, we present some results below using both states as 
model states.

\begin{figure}[!bt]
\includegraphics[angle=270,width=8cm]{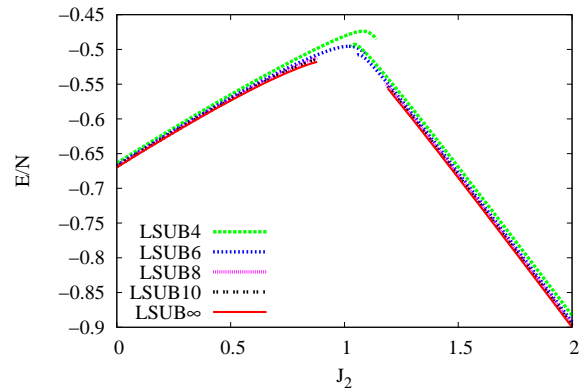}
\caption{CCM results for the gs energy, $E/N$, for the N\'{e}el and
  N\'{e}el$^{\ast}$ states for the spin-$\frac{1}{2}$ $J_{1}$--$J_{2}$ Heisenberg
  antiferromagnet on the checkerboard lattice (with $J_{1}=1$) versus $J_2$.
  The LSUB$m$ approximations for $m=\{4,6,8,10\}$ are shown together with the
  corresponding LSUB$\infty$ extrapolation from using Eq.~(\ref{Extrapo_E}) 
  with $m=\{4,6,8,10\}$.}
\label{E}
\end{figure}

In Fig.~\ref{E} we show both the LSUB$m$ and the extrapolated LSUB$\infty$
results for the gs energy, $E/N$, of both the N\'{e}el and 
N\'{e}el$^{\ast}$  phases.  
\begin{figure*}[!tb]
\mbox{
   \subfloat[N\'{e}el and N\'{e}el$^{\ast}$ states]{\scalebox{0.3}{\includegraphics[angle=270]{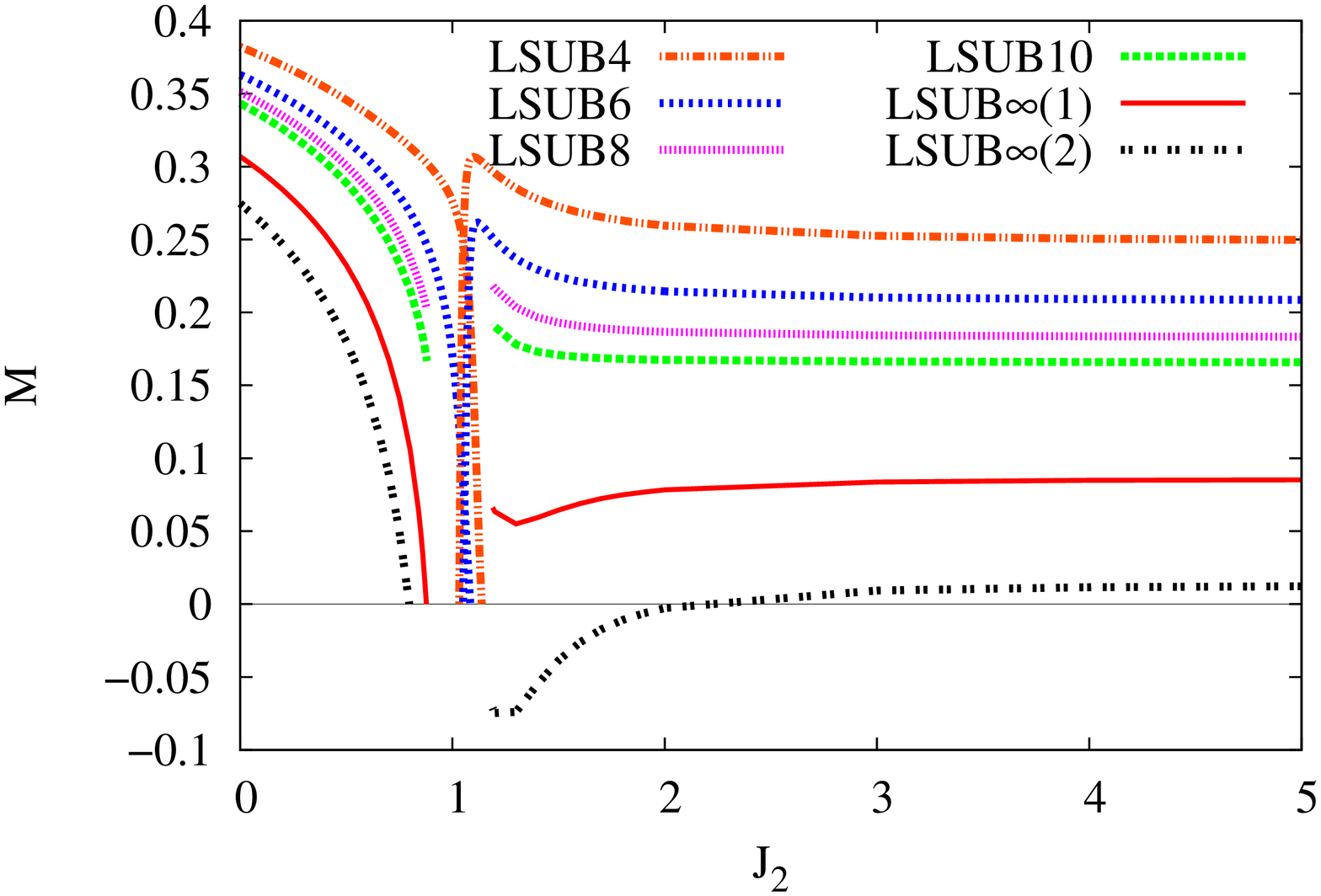}}}
\subfloat[N\'{e}el and striped states]{\scalebox{0.3}{\includegraphics[angle=270]{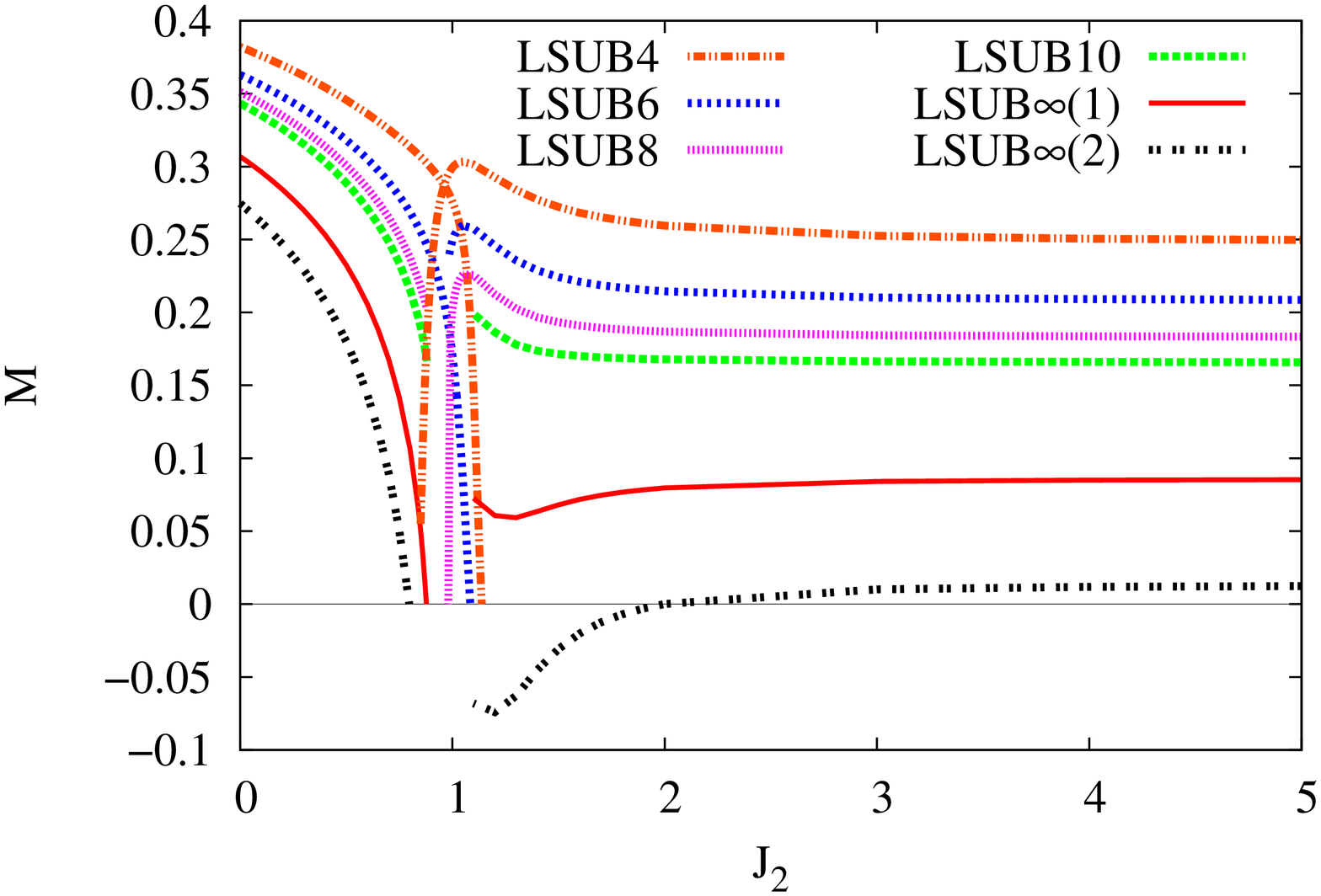}}}
}
\caption{CCM results for the gs magnetic order parameter, $M$, for the N\'{e}el and
  N\'{e}el$^{\ast}$ states for the spin-$\frac{1}{2}$ $J_{1}$--$J_{2}$ Heisenberg
  antiferromagnet on the checkerboard lattice (with $J_{1}=1$) versus $J_2$;
  (a) using the N\'{e}el and N\'{e}el$^{\ast}$ states as model states, and 
  (b) using the N\'{e}el and striped states as model states.  In both cases
  the LSUB$m$ approximations for $m=\{4,6,8,10\}$ are shown together with the
  corresponding LSUB$\infty$(1) and LSUB$\infty$(2) extrapolation from using 
  Eqs.~(\ref{Extrapo_M}) and (\ref{Extrapo_M_Frust}) respectively, with $m=\{4,6,8,10\}$.}
\label{M}
\end{figure*}
As noted briefly above,
we now observe more clearly that each of
the LSUB$m$ energy curves based on a particular model state terminates at some
critical value of $\kappa$ (that itself depends on the LSUB$m$ 
approximation used), beyond which no real CCM solution can be found.  
Since the CCM LSUB$m$ solutions require increasingly more computational power to
obtain (to a given level of numerical accuracy) as the termination
points are approached, it is computationally costly to
determine the actual termination points to a high degree of accuracy.  We
note that in Fig.~\ref{E} results are shown for each LSUB$m$ case down 
to values of $\kappa$ below which for the N\'{e}el$^{\ast}$ phase, and 
up to values of $\kappa$ above which for the N\'{e}el phase, real solutions based on the 
respective model state cease to exist.  As noted above, in all cases the corresponding 
termination point at a given LSUB$m$ level shown in Fig.~\ref{Ediff_ColStriped_stateC}
for the striped state is lower than that for the equivalent N\'{e}el$^{\ast}$ model state case.
 
We note however that, as is usually the case, the CCM LSUB$m$ results for finite $m$ values for
both the N\'{e}el and N\'{e}el$^{\ast}$ phases shown in Fig.~\ref{E} extend beyond the corresponding
LSUB$\infty$ transition points.  For large values of $m$ the LSUB$m$
transition points are quite close to the actual quantum critical points (QCPs)
where that phase ends.  For example, the LSUB10 termination points
shown in Fig.~\ref{E} are at 
$\kappa ^{{\rm N\acute{e}el}}_{t} \approx 0.88$ for the N\'{e}el state
and $\kappa ^{{\rm N\acute{e}el}^{\ast}}_{t} \approx 1.2$ for the N\'{e}el$^\ast$ state.
The CCM results show a clear intermediate regime in which neither of the 
quasiclassical AFM states (N\'{e}el and N\'{e}el$^{\ast}$) is stable.

We now discuss the magnetic order parameter (viz., the average
on-site magnetization), $M$, in order to
investigate the stability of the quasiclassical magnetic LRO.  
Our CCM results for $M$ are shown in Fig.~\ref{M} for each of
the N\'{e}el, N\'{e}el$^{\ast}$, and striped  phases.
Our extrapolated results for $M$ in the N\'{e}el phase are seen 
to be somewhat sensitive to whether we use the scheme of Eq.~(\ref{Extrapo_M}) 
or that of Eq.~(\ref{Extrapo_M_Frust}).
As we have indicated previously the scheme of Eq.~(\ref{Extrapo_M}) is 
appropriate only for small values of $J_2$.  For example, in the
square-lattice limit, $J_{2}=0$, we obtain the extrapolated LSUB$\infty$(1)
result $M \approx 0.3069$ from the use of Eq.~(\ref{Extrapo_M}) and the LSUB$m$
values with $m=\{4,6,8,10\}$.  Very similar values are obtained with
the alternative data sets $m=\{4,6,8\}$ and $m=\{6,8,10\}$.  We 
note that for the square-lattice HAFM no dynamic (or geometric)
frustration exists and the Marshall-Peierls sign 
rule\cite{Marshall-Peierls} applies and may hence be used to
circumvent the QMC ``minus-sign problem.''
The QMC result,\cite{Sandvik:1997} $M=0.3070 \pm 0.0003$,
is thus extremely accurate for this limiting ($J_{2}=0$) case
only.  Our own CCM result using the scheme of Eq.~(\ref{Extrapo_M})  
is thus in excellent agreement with it.  By contrast, the 
extrapolation scheme of Eq.~(\ref{Extrapo_M_Frust}), which is
appropriate for (highly) frustrated systems, gives a much poorer
estimate of $M \approx 0.275$.

The magnetization results show clear evidence for the melting of
N\'{e}el order at a value $\kappa = \kappa_{c_{1}} < \kappa_{{\rm cl}}
= 1$, with results for $\kappa_{c_{1}}$ that are very close to the 
corresponding termination point $\kappa ^{\rm N\acute{e}el}_{t}$ discussed
above.  At such values $\kappa \lesssim 1$, where the system
is highly frustrated, the extrapolation scheme of Eq.~(\ref{Extrapo_M_Frust})
is more appropriate, as we have indicated previously, and 
its use with the LSUB$m$ data set $m=\{4,6,8,10\}$ gives us
our first estimate for the quantum critical point (QCP) at which N\'{e}el order
vanishes, $\kappa_{c_1} \approx 0.796$.  Very similar results are found by using
the alternative data sets $m=\{4,6,8\}$ and $m=\{6,8,10\}$.  Combining
all these results gives the estimate $\kappa_{c_1} \approx 0.80 \pm 0.01$.  (By
contrast, the use of the scheme of Eq.~(\ref{Extrapo_M}), which is 
inappropriate in this frustrated region near a QCP, gives a value  
$\kappa_{c_1} \approx 0.87 \pm 0.01$.)  
  
The results in Fig.~\ref{M}(a) for $M$ using the N\'{e}el$^{\ast}$ state as CCM
model state are seen to be qualitatively very different from those using
the N\'{e}el state as model state.  Indeed, all of the evidence from 
Fig.~\ref{M}(a) is that $M$ is either zero or very close to zero over the
entire range for which CCM extrapolated LSUB$\infty$ solutions exist using 
the N\'{e}el$^{\ast}$ state as model state.  The more appropriate extrapolation scheme
of Eq.~(\ref{Extrapo_M_Frust}) in this regime with $\kappa \gtrsim 1$ gives either negative
values for $M$ or positive values very close to zero
over the entire range shown in Fig.~\ref{M}, while even the inappropriate
scheme of Eq.~(\ref{Extrapo_M}) gives only a very small and almost constant value
of $M \approx 0.08$ over the same range.  Our results in the high-frustration 
regime ($\kappa \gg 1$) using the appropriate scheme of 
Eq.~(\ref{Extrapo_M_Frust}) are given extra credence by the fact that we see
clearly from Fig.~\ref{M} that $M \rightarrow 0$ rather accurately in the large
$\kappa$ limit, which is the exact result for this limit where the model
reduces to unlinked 1D spin-$\frac{1}{2}$ chains.
It is clear that at best the existence of the N\'{e}el$^{\ast}$ phase 
in the spin-$\frac{1}{2}$ case is extremely fragile from this evidence.  More 
likely, it is not the stable gs phase for any value of $\kappa$, based
on the results for $M$.

For this reason we have repeated the calculations for $M$, but now using
the striped state as CCM model state, even though we found it to have
a slightly higher energy for all values of $\kappa$ than that of the 
N\'{e}el$^{\ast}$ state.  Results are shown in Fig.~\ref{M}(b).  
Figures \ref{M}(a) and \ref{M}(b)
show very similar results for $M$ for the striped state and the
N\'{e}el$^{\ast}$ state, except very near their corresponding termination points.
All of the evidence so far is that neither state is the stable gs phase
for any value of $\kappa$.  Since the results for the order parameter $M$ are so
similar for the N\'{e}el$^{\ast}$ state and the striped state, and since
the former has a slightly lower energy, we henceforth restrict ourselves for
larger values of $\kappa$ to use of the N\'{e}el$^{\ast}$ state as
CCM model state.

\begin{figure}[!tb]
\begin{center}
\mbox{
\subfloat{\includegraphics[width=6cm,height=6cm,angle=270]{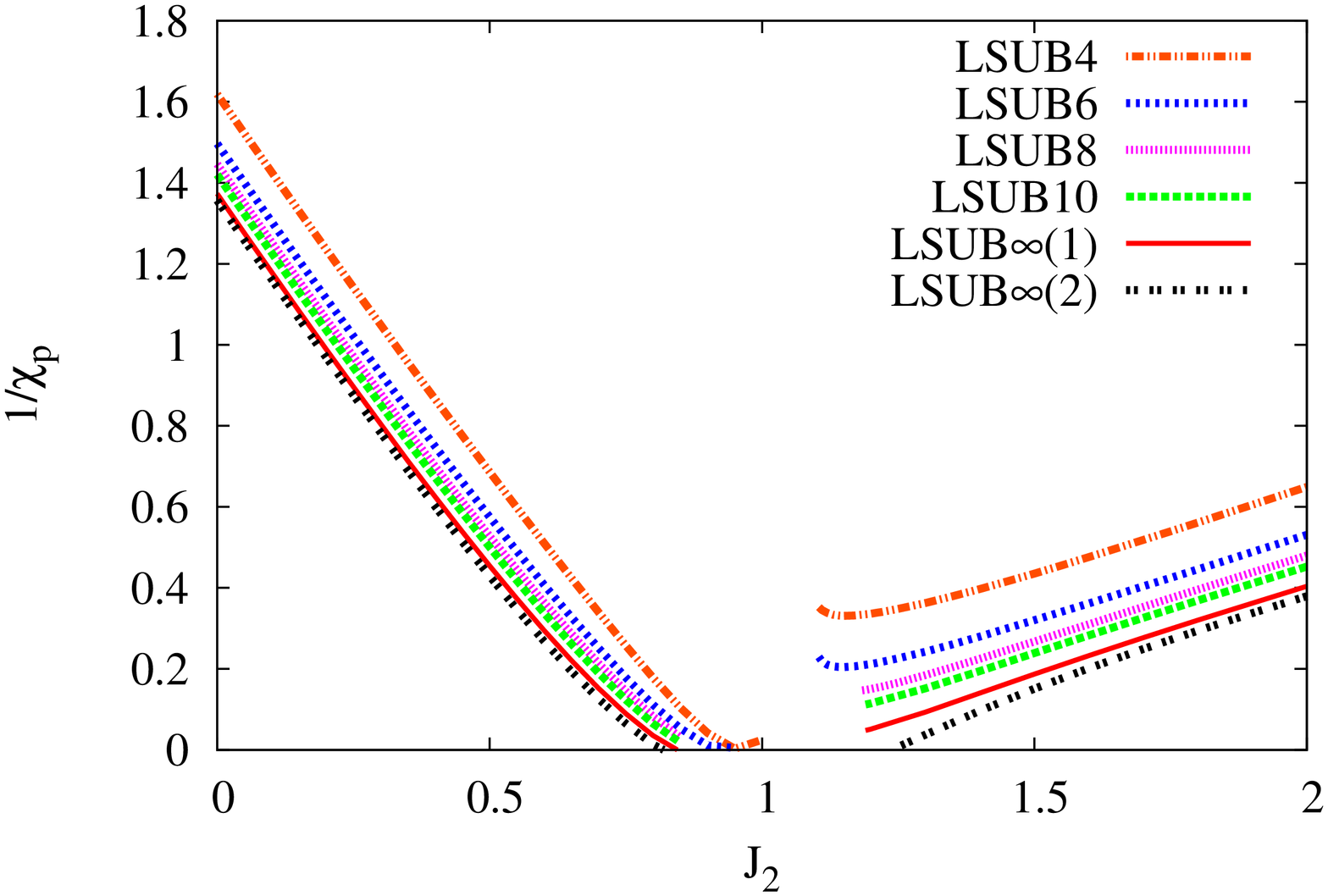}}
\raisebox{-3.5cm}{
\subfloat{\includegraphics[width=2.2cm,height=2.2cm]{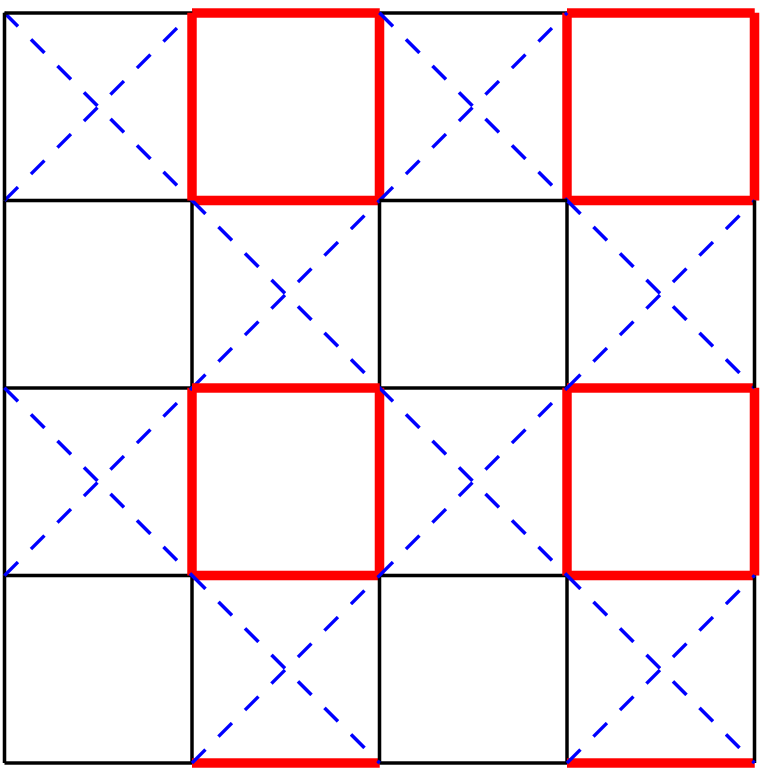}}
}
}
\caption{(Color online) Left: CCM results for the inverse plaquette
  susceptibility, $1/\chi_p$,  versus $J_2$, using the N\'{e}el and
  N\'{e}el$^{\ast}$ states as model states, for the spin-$\frac{1}{2}$ $J_{1}$--$J_{2}$ Heisenberg
  antiferromagnet on the checkerboard lattice (with $J_{1}=1$).
  The LSUB$m$ approximations for $m=\{4,6,8,10\}$ are shown together with the
  corresponding LSUB$\infty$(1) and LSUB$\infty$(2) results from using
  the extrapolations schemes of Eqs.~(\ref{Extrapo_inv-chi-1}) 
  and (\ref{Extrapo_inv-chi-2}) respectively, with $m=\{4,6,8,10\}$.
  Right: The perturbations (fields) $F=\delta\, \hat{O}_p$ for 
  the plaquette susceptibility $\chi_p$.  Thick (red) 
  and thin (black) lines correspond respectively
  to strengthened and weakened NN exchange couplings, where $\hat{O}_p =
  \sum_{\langle i,j \rangle} a_{ij}
  \mathbf{s}_{i}\cdot\mathbf{s}_{j}$, and the sum runs over all NN
  bonds, with $a_{ij}=+1$ and $-1$ for thick (red) and thin (black)
  lines respectively.}
\label{X}
\end{center}
\end{figure}

It is reasonably well established from earlier numerical studies using 
ED\cite{Fouet:2003} and strong-coupling expansion 
techniques\cite{Berg:2003,Brenig:2002,Brenig:2004} that the gs phase
of the spin-$\frac{1}{2}$ HAFM on the checkerboard lattice (i.e., our
model at the isotropic point $J_{2}=J_{1}$) is a plaquette valence-bond
crystal (PVBC) with long-range quadrumer order.  Further evidence for such
a valence-bond solid built from disconnected 4-spin singlets
comes from a ``fermionic'' SU($n$) generalization of the SU(2) group in the
large-$n$ limit.\cite{Canals:2002}  There is broad agreement from all 
this work that the PVBC phase comprises singlet plaquettes on the squares
in Fig.~\ref{model_bonds} without crossed links, as shown in Fig.~\ref{X}.
Thus, in order to get more information on the phase that occurs after the melting
of N\'{e}el order at $\kappa = \kappa_{c_1}$ we now investigate the 
possibility that it might be a PVBC state of the sort shown in Fig.~\ref{X}.

To do so we first consider a
generalized susceptibility $\chi_F$ that describes the response of the
system to a perturbation described by a ``field'' operator $F$.  
A field term $F=\delta\,\hat{O}_F$ is thus added to the Hamiltonian 
of Eq.~(\ref{H}).  The energy per site in a given state,
$E(\delta)/N \equiv e(\delta)$, is then calculated for the perturbed
Hamiltonian $H+F$, and the susceptibility of the
system to the perturbation $F$ is defined as $\chi_{F} \equiv -
\left. (\partial^2{e(\delta)})/(\partial {\delta}^2)
\right|_{\delta=0}$.  An instability of the state against the
perturbation $F$ is signalled by a zero point of $\chi_F^{-1}$ or,
equivalently, by a divergence of $\chi_F$.  In our case
we first use the CCM to calculate $\chi_F$, using a specific model 
state, in various LSUB$m$ approximations.  Although rather less empirical
experience is available for the $m \to \infty$ extrapolation of the CCM
data for $\chi_F$ than for other quantities such as the gs energy $E$ or
the order parameter $M$, we have found previously\cite{Farnell:2011} that
the same extrapolation used for the gs energy [i.e., $\chi_{F}(m) =
d_{0}+d_{1}m^{-2}+d_{2}m^{-4}$] fits the data most accurately, at least
in regions not too close to a divergence of the susceptibility.  We also 
saw previously\cite{Farnell:2011} that a corresponding extrapolation of the 
inverse susceptibility,
\begin{equation}
\chi_{F}^{-1}(m) = x_{0}+x_{1}m^{-2}+x_{2}m^{-4}\,,             \label{Extrapo_inv-chi-1}
\end{equation}
gave very consistent results that agreed very closely with those from
the corresponding above extrapolation of $\chi_{F}$, except again in
regions close to where $\chi_{F}^{-1} \to 0$.  Since, as we will see below,
we will be especially interested in precisely such regions over a large
range of values of $\kappa$, we also use the fitting function,
\begin{equation}
\chi_{F}^{-1}(m) = y_{0}+y_{1}m^{-y_2}\,.            \label{Extrapo_inv-chi-2}
\end{equation}

To calculate the susceptibility, $\chi_p$, of our system against PVBC ordering
we thus set the operator $\hat{O}_F$ to $\hat{O}_p$ as illustrated in the right panel
and the caption of Fig.~\ref{X}.  The perturbation field $F$ thus breaks
the translational symmetry of $H$.  We show CCM results in the left panel
of Fig.~\ref{X} using both the N\'{e}el and N\'{e}el$^{\ast}$ states as 
model states.  We first observe that for smaller values of $J_2$
(i.e., on the N\'{e}el side) the two extrapolations agree very closely, even 
near the point at which $\chi_p^{-1}$ goes to zero.
Thus, the extrapolated inverse plaquette susceptibility using the 
LSUB$m$ data set $m=\{4,6,8,10\}$ ($m=\{6,8,10\}$) vanishes on the N\'eel
side at $\kappa \approx 0.843$ ($\kappa \approx 0.833$) using the extrapolation 
scheme of Eq.~(\ref{Extrapo_inv-chi-1}), and 
at $\kappa \approx 0.820$ ($\kappa \approx 0.775$) using the extrapolation 
scheme of Eq.~(\ref{Extrapo_inv-chi-2}).    
Since the exponent $y_2$ in Eq.~(\ref{Extrapo_inv-chi-2}) falls rather sharply to
a value close to 1 near the point at which $\chi_p^{-1}$ vanishes, the best 
estimate for this point is more likely to come from the extrapolation scheme
of Eq.~(\ref{Extrapo_inv-chi-2}) than from that of  Eq.~(\ref{Extrapo_inv-chi-1}).   

Combining all these results gives our best estimate of
$\kappa \approx 0.79 \pm 0.03$ for the point at which
the  N\'{e}el phase becomes susceptible to PVBC ordering.  This is in very
good agreement with the above estimate of $\kappa_{c_1} \approx 0.80 \pm 0.01$ at which
N\'{e}el LRO disappears as measured by our results for the order parameter $M$.  
Thus, our results show no evidence at all for a coexistence region in which
N\'{e}el and PVBC ordering are both present, such as has been suggested
might occur,\cite{Sachdev:2002} although we cannot exclude the
possibility of a very narrow region of coexistence confined to
the region $0.79 \lesssim \kappa \lesssim 0.81$.
Our findings are in agreement with ED results\cite{Sindzingre:2002} 
for the same spin-$\frac{1}{2}$ anisotropic
planar pyrochlore model that reach the conclusion that, if present at all,
any such coexistence region is very narrow indeed.  As has been discussed in detail
elsewhere,\cite{Starykh:2005} the QCP at $\kappa_{c_1}$ between the 
N\'{e}el and PVBC phases is both forbidden as a continuous transition within
standard Landau-Ginzburg theory and does not seem either to be a viable
candidate for a deconfined (continuous) transition.  The shape of the magnetization curves
in Fig.~\ref{M} on the N\'{e}el side, which show a rapid decrease near $\kappa_{c_1}$,
is perhaps more indicative of a first-order transition, as we have observed 
previously,\cite{Farnell:2011} although such evidence should not 
be regarded as conclusive. 

We also observe from Fig.~\ref{X} that with the N\'{e}el$^{\ast}$ state used as 
our CCM model state the two extrapolations of Eqs.~(\ref{Extrapo_inv-chi-1})  
and (\ref{Extrapo_inv-chi-2}) for the inverse 
plaquette susceptibility, $\chi_p^{-1}$, agree quite closely at larger values
of $\kappa$ but diverge slightly at smaller values, where 
$\chi_p^{-1}$ itself becomes small.  As $\kappa \to \infty$ we observe that
the exponent $y_2$ in Eq.~(\ref{Extrapo_inv-chi-2}) appears to approach the value
1.5 [rather than 2 as in Eq.~(\ref{Extrapo_inv-chi-1})], and then drop to a
value close to 1 as $\chi_p^{-1}$ approaches zero.  For these reasons again, we
expect the extrapolation of Eq.~(\ref{Extrapo_inv-chi-2}) to be more exact,
especially in regions where $\chi_p^{-1}$ becomes small.
Thus, the extrapolated inverse plaquette susceptibility using the 
LSUB$m$ data set $m=\{4,6,8,10\}$ ($m=\{6,8,10\}$) vanishes on the N\'{e}el$^{\ast}$ 
side at $\kappa \approx 1.238$ ($\kappa \approx 1.216$) using the extrapolation 
scheme of Eq.~(\ref{Extrapo_inv-chi-2}).  Although,
as we have seen, the N\'{e}el$^{\ast}$ state does not appear to exist as
as a stable gs phase (since its order parameter $M$ seems to vanish
for all values of $\kappa$, nevertheless the results using it as a CCM
model state provide a robust basis for the calculation of $\chi_p$, and give
an estimate $\kappa_{c_2} \approx 1.22 \pm 0.02$ for the upper QCP at which
PVBC order disappears.

However, we are now led to the question of what is the actual gs phase of the model for 
larger values of frustration, $\kappa > \kappa_{c_2}$, beyond the upper QCP (at 
$\kappa = \kappa_{c_2}$) at which the PVBC phase ceases to exist 
as a stable gs phase.  In the first place it 
is clear that such a QCP must exist since in the limit $\kappa \rightarrow \infty$ one has
the physics of decoupled HAFM 1D chains, which are known to exhibit Luttinger
spin-liquid behavior, and are typified by a gapless excitation spectrum and spin-spin 
correlations that decay algebraically with inter-spin separation distance.  It was 
argued\cite{Starykh:2002} that for large values of $\kappa$, where the 1D
chains are weakly coupled, the gs phase might be a 2D sliding Luttinger liquid phase
(characterized by the absence of LRO and with massless deconfined
spinons as the elementary excitations)
that joined smoothly to the $\kappa \to \infty$ limit.  It was later 
shown,\cite{Starykh:2005} by a more careful analysis of the relevant terms
near the Luttinger liquid fixed point of the independent 1D spin chain, that
this finding was incorrect.  Instead, using techniques that combine
renormalization group ideas and 1D bosonization and current algebra methods,
it was shown that in this large-$\kappa$
regime the gs phase is of spontaneously dimerized type, with a staggered
ordering of dimers along the parallel chains (viz., the diagonals in
Fig.~\ref{model_bonds}).  Such a crossed-dimer valence-bond crystal (CDVBC)
phase, with twofold spontaneous symmetry breaking and no magnetic order, was
independently confirmed\cite{Arlego:2007} by a series expansion (SE) technique
based on the flow equation method.

\begin{figure}[!t]
\begin{center}
\mbox{
\subfloat{\includegraphics[width=6cm,height=6cm,angle=270.]{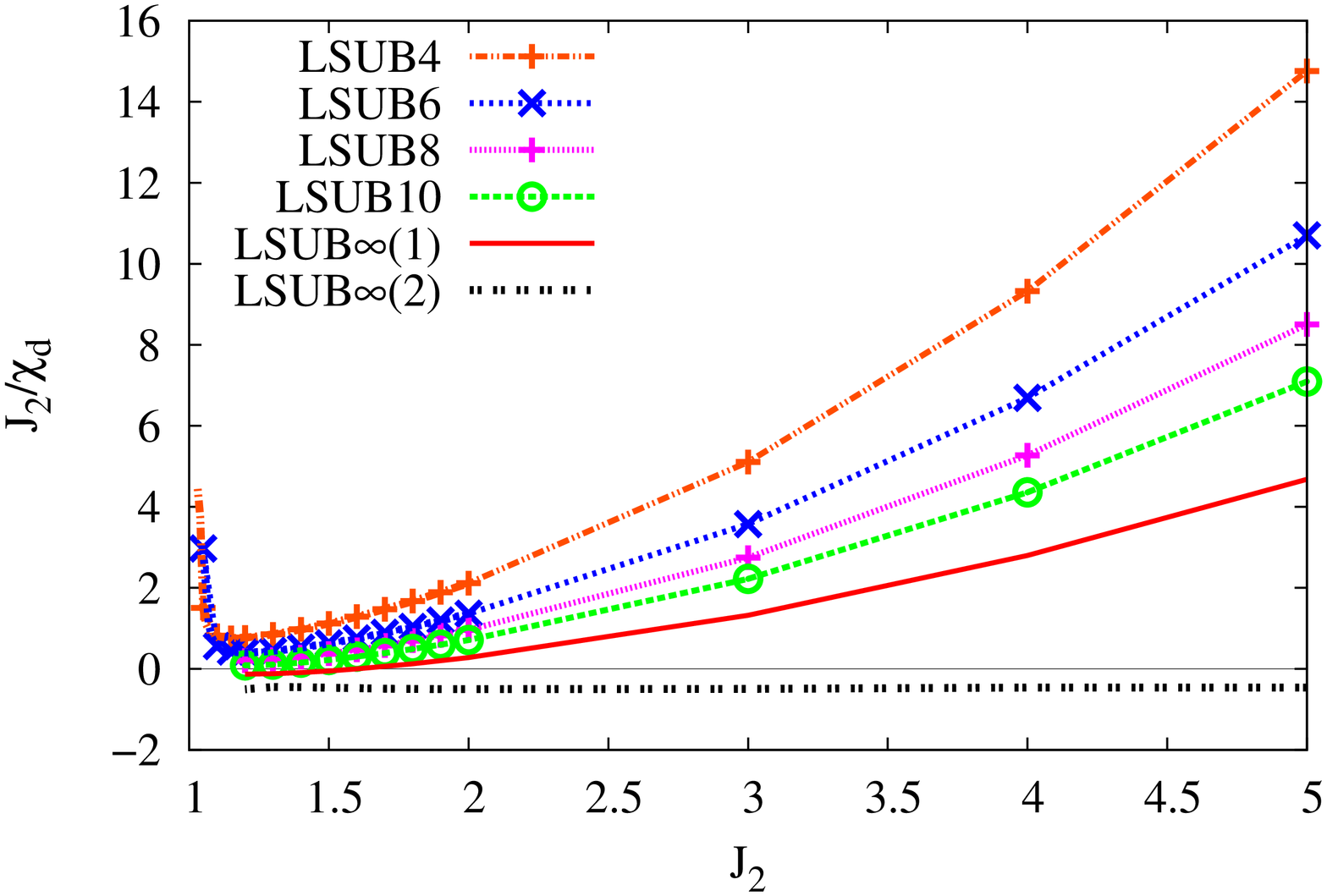}}
\raisebox{-3.5cm}{
\subfloat{\includegraphics[width=2.2cm,height=2.2cm]{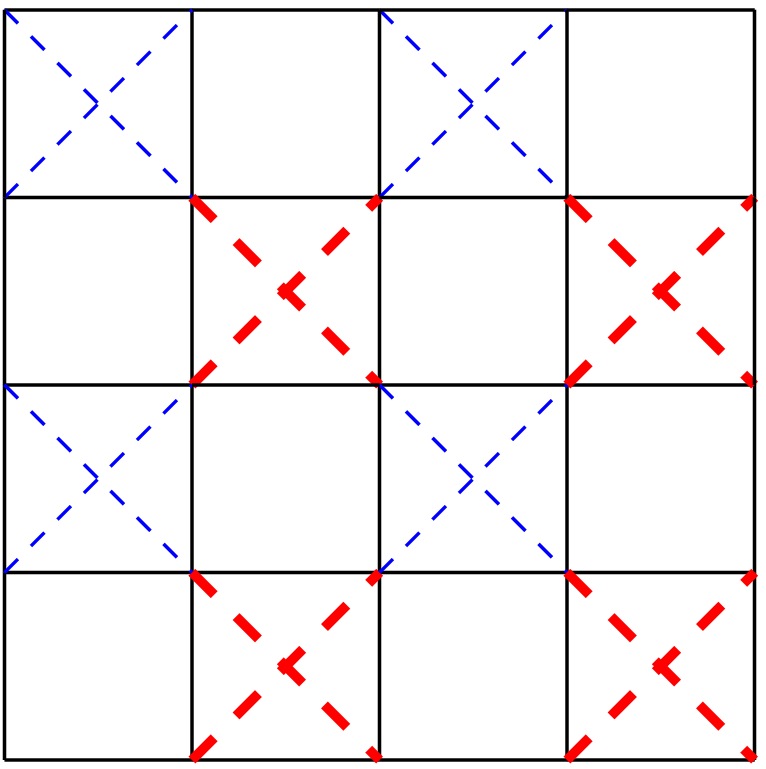}}
}
}
\caption{(Color online) Left: CCM results for the scaled inverse crossed dimer
  susceptibility, $J_{2}/\chi_d$, using the N\'{e}el$^{\ast}$ state as model 
  state, for the spin-$\frac{1}{2}$ $J_{1}$--$J_{2}$ Heisenberg
  antiferromagnet on the checkerboard lattice (with $J_{1}=1$) versus $J_2$.
  The LSUB$m$ approximations for $m=\{4,6,8,10\}$ are shown together with the
  corresponding LSUB$\infty$(1) and LSUB$\infty$(2) results from using 
  the extrapolations schemes of Eqs.~(\ref{Extrapo_inv-chi-1}) 
  and (\ref{Extrapo_inv-chi-2}) respectively, with $m=\{4,6,8,10\}$.
  Right: The perturbations (fields) $F=\delta\, \hat{O}_d$ for 
  the dimer susceptibility $\chi_d$.  Thick (red) dashed
  and thin (blue) dashed lines correspond respectively
  to strengthened and weakened NNN exchange couplings, where $\hat{O}_{d} =
  \sum_{\langle \langle i,k \rangle \rangle} a_{ik}
  \mathbf{s}_{i}\cdot\mathbf{s}_{k}$, and the sum runs over the NNN diagonal
  bonds of the checkerboard lattice, with $a_{ik}=+1$ and $-1$ for thick (red) 
  dashed and thin (blue) dashed lines respectively.}
\label{X_d}
\end{center}
\end{figure}

The CDVBC phase is illustrated in the right panel of Fig.~\ref{X_d}.  Similarly
to what was done above for the PVBC susceptibility, $\chi_p$, we can now calculate 
the susceptibility, $\chi_d$ of our system against CDVBC ordering
by setting the perturbation operator $\hat{O}_F$ to the 
operator $\hat{O}_d$ illustrated in the right panel
and the caption of Fig.~\ref{X_d}.  Since in the large $J_2$ limit the energy
scales linearly with $J_2$ (as may clearly also be observed from 
Figs.~\ref{Ediff_ColStriped_stateC}(a) and \ref{E}, we show our CCM
results in Fig.~\ref{X_d} for the scaled inverse dimer susceptibility,
$J_{2}/\chi_d$ as a function of $J_2$.  

Interestingly, in this case, the extrapolation scheme
of Eq.~(\ref{Extrapo_inv-chi-1}) does not fit the LSUB$m$ data points at all well,
and consequently gives a very poor fit.  The reason becomes quite evident when
the extrapolation scheme of Eq.~(\ref{Extrapo_inv-chi-2}) is used instead.  It
is then observed that that the scaling exponent $y_2$ approaches the value 0.75 
for large values of $J_2$, and even for smaller values near the QCP at
$\kappa=\kappa_{c_2}$ only rises slightly to values that approach 1.  
It is clear that the extrapolated value of $J_{2}/\chi_d$ is small, slightly negative,
and almost constant for all values of $\kappa$ shown, when the more flexible
extrapolation scheme of Eq.~(\ref{Extrapo_inv-chi-2}) is used.  
Our results are consistent with the interpretation that
the inverse dimer susceptibility is zero for all values $\kappa$ shown
in Fig.~\ref{X_d}, namely those where we have real solutions to the equations
pertaining to all of the CCM LSUB$m$ schemes used.  The actual lower termination
point is not easy to determine accurately in this way, as we have 
already mentioned above.  However, it is quite clear from Fig.~\ref{X_d} that
the individual LSUB$m$ curves all have a minimum at a value 
$\kappa \approx 1.20 \pm 0.05$, 
and our results are thus consistent with the interpretation that
the inverse dimer susceptibility is zero for all values $\kappa > \kappa_{c_2}$.
In that scenario there is a QCP between the PVBC and CDVBC phases at
$\kappa = \kappa_{c_2}$, and the CDVBC phase
then persists out to the $\kappa \to \infty$ limit of unlinked 1D chains, which 
is itself a singular point.  

Our results give us essentially no information, however, on the nature of the 
transition at $\kappa_{c_2}$.  Starykh {\it et al.}\cite{Starykh:2005} have 
concluded that a continuous transition between the PVBC and CDVBC phases is
prohibited within the standard Landau-Ginzburg scenario of phase transitions.
They argue further that the most probable alternative is a direct first-order
transition, although a separate possibility is again the existence of an intermediate
coexistence phase with both valence-bond orderings that can then have
separate continuous transitions to both the PVBC and CDVBC phases.  From our results
we clearly favor the former scenario, although we cannot exclude the possibility
of a very narrow such coexistence phase confined to the region $1.20 \lesssim \kappa \lesssim 1.22$.

\section{SUMMARY AND CONCLUSIONS}
\label{summary}
To summarize, we have investigated the gs properties and ($T=0$) phase diagram
of the frustrated spin-$\frac{1}{2}$ antiferromagnetic $J_{1}$--$J_{2}$ model 
(with $J_{2} \equiv \kappa J_{1}\,; J_{1}>0)$ on
the 2D checkerboard lattice, using the CCM carried out to high orders.  

In common with
most other calculations on the model we find that the gs phase is an AFM 
N\'{e}el-ordered state for $\kappa < \kappa_{c_1}$, at which
point the staggered N\'{e}el magnetization vanishes.  Our best estimate for
this lower QCP is $\kappa_{c_1} \approx 0.80 \pm 0.01$.  This is in reasonable 
agreement, but probably more accurate than, a corresponding 
estimate of $\kappa_{c_1} \approx 0.75$ from an ED 
study\cite{Sindzingre:2002} on samples of $N=16,32,36$ spins.  Since we calculate
that the N\'{e}el-ordered state becomes susceptible to PVBC ordering
at $\kappa \approx 0.79 \pm 0.03$ our results
point then to a direct transition from the N\'{e}el-ordered gs phase to
a PVBC phase at $\kappa = \kappa_{c_1}$, although we cannot exclude entirely
the small possibility of a vey narrow coexistence region.  This finding is in good 
agreement with that found from ED studies.\cite{Sindzingre:2002}  We estimate
that any such coexistence region of AFM N\'{e}el ordering and quadrumer plaquette
ordering is confined to the parameter range $0.79 \lesssim \kappa \lesssim 0.81$.
Our estimate for the lower boundary at which the quadrumer PVBC ordering vanishes
is in excellent agreement with the corresponding value $\kappa \approx 0.80 \pm 0.01$ 
from a high order SE calculation\cite{Brenig:2004} that starts from the limit of 
uncoupled quadrumers.  A recent tensor network simulation of the model\cite{Chan:2011}
gives a somewhat higher value of $\kappa_{c_1} \approx 0.88$ for the QCP from
the N\'{e}el gs phase to the PVBC gs phase.

From our CCM calculations we estimate that the quadrumer order of the PVBC state 
vanishes at a higher QCP at $\kappa = \kappa_{c_2} \approx 1.22 \pm 0.02$.  This is
in reasonable agreement, although somewhat higher than the corresponding estimates 
$\kappa \approx 1.095 \pm 0.035$ from a high-order SE calculation,\cite{Brenig:2004}
and $\kappa \approx 1.11$ from a tensor network simulation.\cite{Chan:2011}

Although the striped and N\'{e}el$^{\ast}$ states, which are the fourfold-degenerate 
quasiclassical gs phases at $O(1/s)$ in an expansion in powers of $1/s$ for 
$\kappa > 1$,\cite{Tchernyshyov:2003} provide excellent model states for CCM
calculations at larger values of $\kappa \gtrsim 1$ in the sense of providing 
well-converged LSUB$m$ results, we find that they are not stable gs phases for 
the spin-$\frac{1}{2}$ model for any value of $\kappa$.  This is in sharp 
disagreement with the finding from a recent tensor network simulation\cite{Chan:2011}
that the striped and/or N\'{e}el$^{\ast}$ states form the stable gs phase for all
values $\kappa > \kappa_{c_2}$.  By contrast we find from our CCM calculations
that the N\'{e}el$^{\ast}$ state is susceptible to the formation of a CDVBC
phase (with an inverse susceptibility that is essentially zero) for all values 
$\kappa \gtrsim 1.20 \pm 0.05$.  We conclude that the CDVBC state is thus likely
to be the stable gs phase for $\kappa > \kappa_{c_2}$, although we cannot
exclude the possibility of a very narrow coexistence regime confined to 
the range $1.20 \lesssim \kappa \lesssim 1.22$.  We find no evidence at all
for any region separating the PVBC and CDVBC phases where the N\'{e}el$^{\ast}$ state
would form a stable gs phase, thereby ruling out one of the two scenarios 
postulated by Starykh {\it et al.},\cite{Starykh:2005} although now in complete
agreement with their other scenario for the gs phase diagram, on which we now
provide accurate numerical results for the two QCP's at $\kappa_{c_1}$ and
$\kappa_{c_2}$.

Finally we note that the CCM has been used here with the model or reference states
chosen as classical states built from independent-spin product states.  For larger
values of the frustration parameter, $\kappa > \kappa_{c_1}$, 
we remark that these model states are able (with appropriate choice of additional 
perturbative terms in the Hamiltonian), to describe the susceptibilities of 
the system to form the true gs phasaes perfectly well, 
even though we have shown that the respective extrapolated order 
parameters with respect to these model states 
are essentially zero over the entire range.  Nevertheless, it
would be worthwhile to repeat the CCM calculations directly with dimer and plaquette 
valence-bond states.  Indeed, just such a general CCM approach\cite{Farnell_VBgroundstates:2009} 
has recently been described, which combines exact solutions 
for dimer or plaquette valence-bond solid ground states
with the computational implementation described here\cite{ccm} that is based on
independent-spin product model states.  It would be interesting to use this formalism 
for the present model in order to confirm our results.

\section*{ACKNOWLEDGMENT}
We thank the University of Minnesota Supercomputing Institute for the grant of
supercomputing facilities.

\end{document}